# A Variational Approach for Minimizing Lennard-Jones Energies

Carsten Peterson[1], Ola Sommelius[2] and Bo Söderberg[3]

Department of Theoretical Physics, University of Lund
Sölvegatan 14A, S-223 62 Lund, Sweden

**Abstract:**

A variational method for computing conformational properties of molecules with Lennard-Jones potentials for the monomer-monomer interactions is presented. The approach is tailored to deal with angular degrees of freedom, *rotors*, and consists in the iterative solution of a set of deterministic equations with annealing in temperature.

The singular short-distance behaviour of the Lennard-Jones potential is adiabatically switched on in order to obtain stable convergence.

As testbeds for the approach two distinct ensembles of molecules are used, characterized by a roughly dense-packed ore a more elongated ground state. For the latter, problems are generated from natural frequencies of occurrence of amino acids and phenomenologically determined potential parameters; they seem to represent less disorder than was previously assumed in synthetic protein studies.

For the dense-packed problems in particular, the variational algorithm clearly outperforms a gradient descent method in terms of minimal energies. Although it cannot compete with a careful simulating annealing algorithm, the variational approach requires only a tiny fraction of the computer time.

Issues and results when applying the method to polyelectrolytes at a finite temperature are also briefly discussed.

---
[1] carsten@thep.lu.se
[2] ola@thep.lu.se
[3] bs@thep.lu.se

# 1 Introduction

The determination of configurations of long molecular chains often is a difficult problem. For polyelectrolytes, consisting of identical charged monomers interacting with Coulomb repulsion forces, the ground state is trivial – the monomers form a straight line. The challenge here lies in predicting statistical quantities for finite temperature configurations; thus, the thermodynamics of the system is crucial. For proteins, modelled as a sequence of point-like amino acids interacting with effective pair potentials having local minima, the situation is somewhat different. Here the groundstate is nontrivial; the energy landscape is typically plagued with many local minima. This is the main difficulty here, while the finite temperature properties often are just considered as minor perturbations around the ground state.

Optimization problems with many local minima as in the the protein case are notoriously difficult, and elaborate methods to search the phase space efficiently have been devised. One such method is Simulated Annealing (**SA**) [1], where noise is introduced to emulate a Boltzmann distribution; this enables the system to escape from local minima. Unfortunately, this procedure is quite CPU demanding. The formal temperature in such an approach is merely an artificial parameter; it need not be identified with the physical temperature of the system.

For some optimization problems it has turned out profitable to abandon stochastic methods like SA in favour of deterministic approaches based on the iterative solution of equations originating from a variational scheme. Successful results with such approaches were reported for combinatorial optimization problems that can be mapped onto spin systems [2, 3].

Variational methods have recently also been used to compute configurational properties of polyelectrolytes. In refs. [4, 5], harmonic trial potentials were used with widths and positions as variational parameters. This approach works well as long as the potentials to be approximated have a milder divergence than $1/r^3$. In the Coulomb case (screened or unscreened), which was treated in refs. [4, 5], this requirement is fulfilled. However, for the strongly diverging ($1/r^{12}$) Lennard-Jones (**LJ**) potential, occuring in protein models, a harmonic Ansatz will lead to divergent integrals, as will indeed any smooth Ansatz.

This problem could be overcome in two different ways: In principle, one could use a modified Ansatz for the trial potential, that leads to finite integrals. However, this is difficult to achieve while maintaining the computational simplicity needed for a competitive algorithm. Alternatively, one could keep the harmonic Ansatz, and instead modify the LJ-potential to make it less singular.

The approach of refs [4, 5] applies only to molecules with flexible bonds, however; in this paper we will consider a slightly simplified model with *rigid bonds*, and the dynamics hence limited to the angular degrees of freedom. This simplification is motivated by the assumption that the flexibility of bonds is of minor importance for obtaining an accurate spatial structure. Indeed, several calculations using real chains of amino acids focus entirely on the angular degrees of freedom [6].

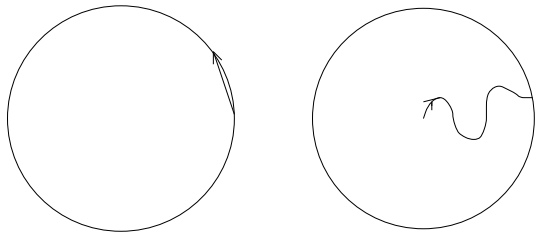

Figure 1: **(a)**. Elementary moves on the unit sphere. **(b)**. Evolution of a MF rotor initialized close to the center, for $T < T_c$.

Thus, we are facing a minimization problem in terms of an energy-function of angular variables. These will be represented by a set of unit vectors, **rotors**, which can be seen as continuous generalizations of discrete Ising spins. In algorithms like gradient descent (**GD**) and SA elementary moves are made with the rotors *on-shell*, ie. restricted to the unit sphere (see fig. 1a).

In ref. [10] the variational mean-field approach, commonly used for discrete spin-systems, was generalized by introducing *mean field rotors*; these can explore the interior of the sphere and can be interpreted as thermal averages of on-shell rotors. The method was successfully explored on the minimal energy problem for charges on a sphere. The corresponding mean field equations are iteratively solved as the temperature $T$ is lowered. Above a critical temperature $T_c$ the system relaxes to a fixed point with all rotors at the center of the sphere. As the temperature is lowered the rotors approach the "shell" (see fig. 1b). Thus, in the variational approach, rather than fully or partly exploring the configuration space, the variables "feel" their ways off-shell in a fuzzy manner towards good on-shell solutions.

The main purpose of this paper is to apply a similar technique to proteins, modelled as a chain of monomers connected by rigid bonds and interacting via LJ-potentials, with the aim of finding the ground state.

Any updating algorithm faces problems with the steepness of the LJ-potential at short distances. We have developed an adiabatic regularization procedure that efficiently handles the short-distance problem, which is useful for any method, not just the variational rotor approach.

Computational simplicity is gained, at the price of sacrificing uniqueness and physical interpretability at non-zero $T$, by approximating expectation values according to

$$\langle E(\cdot) \rangle \to E(\langle \cdot \rangle) \tag{1}$$

The resulting algorithm is shown to yield a GD algorithm in the $T \to 0$ limit.

As LJ-potential testbeds we do not consider real-world proteins. For the purpose of algorithmic development and studies it suffices to study synthetic systems. We have chosen to study two extremes, by considering systems with a dense-packed or elongated ground state, respectively. The latter were generated from empirical independent amino-acid probabilities and phenomenologically determined pairwise forces. As a by-product, it turns out that the resulting LJ-parameters give rise to far less disordered systems than has been assumed in some recent generic investigations of spectra and stability issues [8, 9].

For these testbeds the variational rotor approach is explored for problem sizes $N$ ranging from 10 to 40. For the dense-packed model the variational approach compares favourably with GD with respect to reaching low energy states, while for the elongated model the corresponding gain is smaller. For both models the CPU time consumption is much lower than for SA.

For completeness, we apply the variational rotor approach also to a polyelectrolyte (Coulomb) problem at $T \neq 0$, and discuss the limitations of the approach here.

This paper is organized as follows. In Sect. 2 the variational (mean field) formalism is derived. The generation of synthetic proteins and their LJ-potential couplings are discussed in Sect. 3. The regularization of the LJ-potential is treated in Sect. 4. In Sect. 5 we present the numerical procedures and results for the LJ-potential, and Sect. 6 contains the polyelectrolyte application. A brief summary and outlook can be found in Sect. 7.

## 2 The Variational Approach

### 2.1 Proper Variational Approach

Limiting ourselves to angular degrees of freedom, we consider an energy function

$$E = E(\mathbf{s}_1, \ldots, \mathbf{s}_N) \tag{2}$$

to be minimized w.r.t. the directions of a set of $N$ distinct $D$-dimensional unit vectors $\mathbf{s}_i$ (*rotors*)

$$|\mathbf{s}_i| = 1 . \tag{3}$$

In instances where the energy landscape contains many local minima, one would typically employ a stochastic technique like SA. In [10] a *mean field* method was developed and numerically explored for the problem of placing charges on a sphere. We will here generalize this technique, and apply it to the case of a protein model with LJ pair-potentials.

For spin systems the mean field approximation can be derived in (at least) three conceptionally distinct ways; from a *variational principle*, from a *saddlepoint approximation*, or using an *intuitive physical argument*. We will here briefly discuss the variational derivation; for the saddle-point approach we refer the reader to e.g. ref. [10].

The variational approach is based on an effective energy Ansatz, $E_V$, which is linear in the spins $\mathbf{s}_i$,
$$E_V(\mathbf{s}_1, \ldots, \mathbf{s}_N) = -\sum_i \mathbf{u}_i \cdot \mathbf{s}_i \tag{4}$$
where the (real, unconstrained) coefficient vectors $\mathbf{u}_i$ are to be considered as *variational parameters*.

Based on the corresponding Boltzmann distribution ($\propto \exp(-E_V/T)$), a free energy, $F_V$, can be defined,
$$F_V(\mathbf{u}_1, \ldots, \mathbf{u}_N) = \langle E \rangle_V - ST, \tag{5}$$
where $S$ is the entropy. $F_V$ is bounded from below by the true equilibrium free energy, $F$, based on the proper Boltzmann distribution $\propto \exp(-E/T)$.

The variational free energy $F_V$ can be written as
$$F_V = -T \log Z_V + \langle E - E_V \rangle_V \tag{6}$$
where $Z_V$ is the variational partition function, while $\langle \rangle_V$ refers to averages w.r.t. the variational Boltzmann distribution. By minimizing $F_V$ with respect to the parameters $\mathbf{u}_i$, the variational equations result. These are best expressed in terms of the *mean fields*
$$\mathbf{v}_i = \langle \mathbf{s}_i \rangle_V \tag{7}$$
which approximate the exact averages $\langle \mathbf{s}_i \rangle$. The mean fields are simple functions of the coefficients $\mathbf{u}_i$,
$$\mathbf{v}_i = \frac{\mathbf{u}_i}{|\mathbf{u}_i|} g(|\mathbf{u}_i|/T) = \mathbf{g}(\mathbf{u}_i/T) \tag{8}$$
where for the case of three dimensions $g$ is given by $g(x) = \coth(x) - 1/x$; its graph is shown in fig. 2. Note that $g(0) = 0$, and that $g(x) \to 1$, when $x \to \infty$; this generic qualitative behaviour holds for any number of dimensions, and implies in particular that as $T \to 0$ the MF rotors $\{\mathbf{v}_i\}$ go on-shell, and can be identified with a low-energy configuration $\{\mathbf{s}_i\}$.

The variation of the entropy term $ST$ yields
$$d(ST) = -\sum_i \mathbf{u}_i d\mathbf{v}_i \tag{9}$$

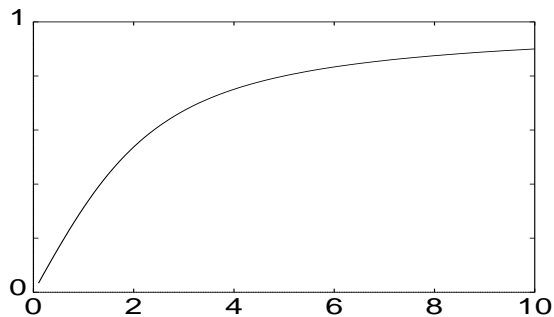

Figure 2: The function $g(x)$ for the case of three dimensions.

and we obtain the variational equations in the form

$$\mathbf{v}_i = \mathbf{g}\left(-\nabla_{\mathbf{v}_i}\langle E\rangle_V/T\right) \qquad (10)$$

## 2.2 Modified Variational Approach

For a strongly singular potential, like LJ, things are complicated by the fact that $\langle E\rangle_V$ is a diverging integral. In other cases, the corresponding integral may be convergent, but difficult to evaluate. These difficulties can be remedied by making the apparently crude replacement

$$\langle E(\mathbf{s}_1,\ldots,\mathbf{s}_N)\rangle_V \to E(\langle \mathbf{s}_1\rangle_V,\ldots,\langle \mathbf{s}_N\rangle_V) = E(\mathbf{v}_1,\ldots,\mathbf{v}_N) \qquad (11)$$

in the expression for $F_V$. This is justified as long as we are only interested in the ground-state, which dominates for $T \to 0$ where the fluctuations vanish and the above approximation becomes exact. For a finite $T$, the replacement in eq. (11) is more questionable, and its use must be justified by other means.

Minimization of the thus modified $F_V$ with respect to the trial parameters $\mathbf{u}_i$ (or $\mathbf{v}_i$) yields a modified set of equations:

$$\mathbf{v}_i = \mathbf{g}\left(-\nabla_i E(\mathbf{v}_1,\ldots,\mathbf{v}_N)/T\right) \qquad (12)$$

In principle, these could be iterated to find a (local) minimum of $F_V$.

However, a question of uniqueness arises: since $E$ really is defined only on-shell, i.e. for $|\mathbf{s}_i| = 1$, off-shell values are not uniquely defined. By adding eg. a term $\propto \mathbf{v}_i^2 - 1$ which is zero on-shell, we can alter the off-shell behaviour of $E$, and hence the solutions to eq. (12). Thus, these make sense only in the $T \to 0$ limit, where the $\mathbf{v}_i$ are forced on-shell, due to the asymptotic behaviour of $g()$.

Another problem is a possible lack of stability of the iterative dynamics of eq. (12). This can be remedied by formally adding a stabilizing term to the energy of the form $-\beta/2 \sum_i \mathbf{v}_i^2$, which is just a constant on-shell.

We then obtain a modified set of variational equations:

$$\mathbf{u}_i = \beta \mathbf{v}_i - \nabla_i E(\mathbf{v}_1, \ldots, \mathbf{v}_N) \tag{13}$$

$$\mathbf{v}_i = \mathbf{g}\left(\frac{\mathbf{u}_i}{T}\right) \tag{14}$$

to be iterated. For $T \to 0$ this turns into a kind of an (on-shell) gradient descent with $\beta$ corresponding to a reciprocal step size. This of course could have been obtained in a much simpler way. However, we aim at an annealing approach: start iterating with a high $T$, for which the modified $F_V$ typically is minimized by a fixpoint with $\mathbf{v}_i \approx 0$. Then keep iterating while slowly letting $T \to 0$. The value of $\beta$ should be chosen to stabilize the iterative dynamics. The idea is that with the soft high $T$ dynamics, the $\mathbf{v}_i$ are allowed to short-cut through an interpolating (off-shell) space. As $T$ slowly falls to zero, the $\mathbf{v}_i$ are eventually forced on-shell, and finally a local minimum of the on-shell energy function $E(\mathbf{s}_i)$ crystallizes out. This way, one can hope to obtain better minima then by just using gradient descent.

## 3 The Lennard-Jones Potential

In models of proteins one often uses a potential between the individual atoms consisting in a sum of "bonded" interactions and "non-bonded" ones. The latter consist of Coulomb and LJ interactions between all the atoms.

It has been argued [12] that a reasonable simplification results from considering the amino acids as *effective monomers*, interacting via *effective potentials*. We will focus on effective LJ potentials. Crucial physical properties like hydrophobicity then are efficiently incorporated by a suitable choice of LJ parameters.

The LJ potential between two monomers $k$ and $l$ is given by

$$V_{kl}(r) = \frac{R_{kl}}{r^{12}} - \frac{A_{kl}}{r^6} \tag{15}$$

An example is shown in fig. 3. It is short-range, and has a very strong (and unphysical) short-distance repulsion. If $A_{kl} > 0$ it has a local minimum located at

$$r_{kl}^0 = \left(\frac{2R_{kl}}{A_{kl}}\right)^{1/6} \tag{16}$$

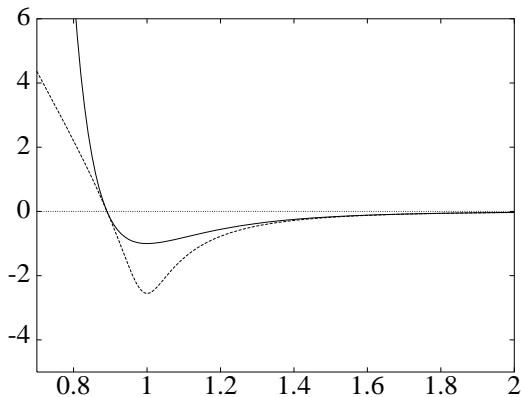

Figure 3: The Lennard-Jones potential (eq. (15)) and a regularized version (eq. (18)) as full and dashed lines, respectively ($R=1$; $A=2$).

The energy to be minimized is given by the sum of all pair-potentials

$$E = \sum_{kl} V_{kl}(r_{kl}) \qquad (17)$$

## 3.1 Regularizing the r → 0 Behaviour

The very steep short-distance behaviour of the LJ-potential might cause problems for updating in minimization algorithms – not only for our variational approach but also for e.g. some Monte Carlo (**MC**) algorithms. A convenient way of controlling the short-distance singularity is by using a modified potential (see fig. 3)

$$\tilde{V}_{kl}(r) = V_{kl}^0 + \frac{1}{\gamma} \log\left(1 + \gamma(V_{kl}(r) - V_{kl}^0)\right) \qquad (18)$$

where $V_{kl}^0$ is the minimum value of $V_{kl}(r)$. The modified potential only has a logarithmic singularity at the origin, and yields the same result in the limit $\gamma \to 0$ as $V$. Furthermore, the position of the minimum is preserved for all values of $\gamma$. The idea is to start with $\gamma > 0$ and then gradually decrease $\gamma$ to 0. In minimization schemes of annealing type, where noise is introduced via a temperature $T$, the decrease in $\gamma$ should be coupled to the annealing in $T$ (see Sect. 4).

## 3.2 Coupling Distributions

We choose the LJ parameters $R_{kl}$ and $A_{kl}$ in two different ways; (A) identical values for all monomers, corresponding to dense packing, and (B) using empirical couplings for a random sequence of aminoacids, based on empirical frequencies of occurrence:

**A** For all pairs, set $R_{kl} = 1$ and $A_{kl} = 2$, leading to $r^0_{kl} = 1$. This choice implies an approximately dense-packed ground state.

**B** Choose $R_{kl}$ and $A_{kl}$ in a phenomenological way as follows. Generate a random sequence of amino-acids according to the known frequencies of occurrence in proteins [13]. The various values for $R_{kl}$ and $A_{kl}$ are derived from the empirical effective LJ couplings given in [12]. The resulting $A_{kl}$ distribution is shown in fig. 4. We note that while this distribution exhibits quite some width, it is far from being as dramatic as in the toy models proposed in refs. [8, 9], where $R_{kl} = 1$ and $A_{kl} = 3.8 + 6.0\,\text{rand}[0,1]$ were used.

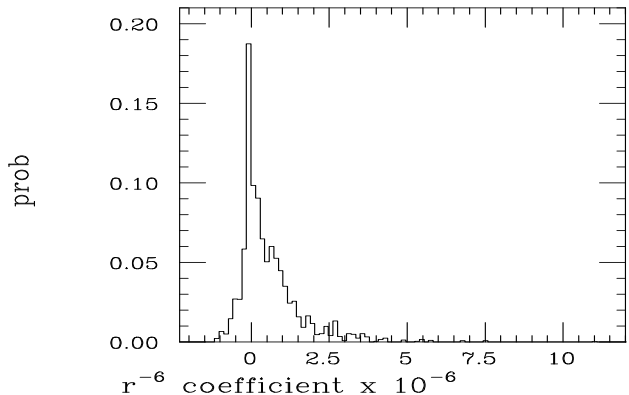

Figure 4: Distribution of $A_{kl}$ according to option **B**.

## 4 The Variational Rotor Algorithm

We next briefly describe how we implement eqs. (13,14) for the regularized LJ-potential, eqs. (15,18). The non-uniqueness due to eq. (11), and the existence of unique pairwise local minima $r^0_{kl}$ in eq. (16), suggests a modified distance expression:

$$\tilde{r}^2_{kl} = \left(\sum_{i\in\sigma(kl)} \mathbf{v}_i\right)^2 + \frac{(r^0_{kl})^2}{|k-l|} \sum_{i\in\sigma(kl)} \left(1 - \mathbf{v}^2_i\right) \qquad (19)$$

where $\sigma(kl)$ denotes the set of bonds connecting monomers $k$ and $l$. To avoid instability in the dynamics, $r^0_{kl}$ is maximized to $\sqrt{|l-k|}$; this value is also used for a negative $A_{kl}$. On-shell, this yields the correct distance, while at the beginning of the simulation when $\mathbf{v}_i = 0$, most pairs are formally at their distance of minimum energy. The regulator $\gamma$ is chosen to depend on $R_{kl}$ as

$$\gamma_{kl} = \frac{4\hat{\gamma}}{R_{kl}} \qquad (20)$$

where $\hat{\gamma}$ is the same for all pairs; this leads to correct relative normalization of the regularized potentials at short distances.

The resulting variational equations (13,14) read

$$\mathbf{u}_i = \beta \mathbf{v}_i + 6 \sum_{\sigma \ni i} \frac{(2R_\sigma - A_\sigma \tilde{r}_\sigma^6) \left( \sum_{j \in \sigma} \mathbf{v}_j - \frac{(r_\sigma^0)^2}{|k-l|} \mathbf{v}_i \right)}{\tilde{r}_\sigma^2 \left( \tilde{r}_\sigma^{12} + \hat{\gamma} \left( 2 - \frac{A_\sigma}{R_\sigma} \tilde{r}_\sigma^6 \right)^2 \right)} \qquad (21)$$

$$\mathbf{v}_i = \mathbf{g}(\mathbf{u}_i/T) \qquad (22)$$

where $\sigma$ denotes a connected subset of the bonds, that includes bond $i$.

The variational algorithm then takes the following form:

1. Initialize:

    1.1 Set $T=T_0$ and $\hat{\gamma}=\hat{\gamma}_0$.

    1.2 Initialize $\mathbf{v}_i$ to small ($\sim .01$) random vectors.

2. Repeat until $\frac{1}{N-1} \sum \mathbf{v}_i^2 > 0.99999$:

    2.1 Update all $\mathbf{v}_i$'s according to eqs. (21) and (22).

    2.2 Anneal: $T = k_T T; \hat{\gamma} = k_\gamma \hat{\gamma}$

The choice of parameter values are shown in table 1; for the stabilizer $\beta$ the choice is motivated by the empirical observation that the $\beta$-values required for high quality solutions scale linearly with $N$. In the gradient descent limit, where $T = 0$ and the rotors are on-shell, $\mathbf{v}_i = \mathbf{u}_i/|\mathbf{u}_i|$, $\beta$ plays the role of an inverse step length.

|  | $\beta$ | $\hat{\gamma}_0$ | $k_\gamma$ | $T_0$ | $k_T$ |
|---|---|---|---|---|---|
| dense | $25N$ | 0.5 | 0.99 | $\beta/3$ | 0.995 |
| real | $100N$ | 0.5 | 0.99 | $\beta/3$ | 0.995 |

Table 1: Parameter settings for the variational algorithm (the first three apply also to GD).

# 5  Numerical Results for the LJ-Potential

We have compared the performance of the variational rotor algorithm to that of (i) a GD algorithm and (ii) a SA method – both for the model A, in which the final structure will be close to a dense-packed configuration, and for the phenomenological model B. The GD method was chosen as the $T = 0$ limit of the variational algorithm, with identical parameter settings for $\beta$, $\hat{\gamma}_0$ and $k_\gamma$ (cf. Table 1); thus, the regularized LJ potentials were used also here. The SA simulations were made using Metropolis updating, with an initial temperature $T_0$ given by three times the modulus of the lowest energy achieved by the variational method in case A (for simplicity, the same $T_0$ was used also for the B problems of the same size). With an annealing rate per sweep of $k_T = 0.99999$, approximately $2.5 \cdot 10^6$ sweeps were required, with a sweep defined by $N$ attempted single-rotor updates.

These choices of parameters and number of sweeps were based on reasonable trade-offs between solution quality and consumption of CPU time.

|      | CPU-time |       |        |
| ---- | -------- | ----- | ------ |
|      | N = 10   | 20    | 40     |
| GD   | 12       | 70    | 700    |
| Var. | 15       | 120   | 1500   |
| MC   | 4000     | 25000 | 200000 |

Table 2: Approximate CPU consumption for model B in seconds on a DEC Alpha workstation for the different approaches. Similar numbers hold for model A.

The reference values for the ground-state energies thus obtained by SA, at the price of very high CPU consumption (see table 2), are only rarely equalled by the energies obtained using the two deterministic approaches. In model A (dense-packed) the variational algorithm clearly outperforms the GD one (see table 3). However, for the slightly more realistic model B the two approaches seem comparable. A possible explanation for this is that the main advantage of the variational approach is the extra degrees of freedom since the rotors can go off-shell; this facilitates the escape from local minima, which probably is more important in the dense packed case.

|   | N  | $\langle E_{GD} - E_{var} \rangle$ | ± | $\langle E_{var} \rangle - E_{SA}$ | ± | # probl. | # runs | # sweeps Var. | GD |
|---|----|------|-----|------|-----|----|-----|-------|-------|
|   | 10 | 0.8  | 0.1 | 1.0  | 0.1 | 1  | 100 | 2200  | 550   |
| A | 20 | 3.0  | 0.4 | 4.8  | 0.2 | 1  | 100 | 2300  | 1300  |
|   | 40 | 7.7  | 0.8 | 6.4  | 0.3 | 1  | 100 | 3000  | 4400  |
|   | 10 | 0.6  | 0.5 | 6.0  | 0.9 | 10 | 10  | 4100  | 3800  |
| B | 20 | 2.3  | 1.2 | 14.9 | 1.8 | 10 | 10  | 9200  | 10400 |
|   | 40 | 9.1  | 4.3 | 59.7 | 6.1 | 5  | 10  | 16100 | 15300 |

Table 3: Average performance differences. The number of runs refers to the variational and GD approaches. For each problem a single run was performed with SA.

# 6 Polyelectrolytes

In this section we briefly discuss how the variational rotor approach can be applied to the case of a polyelectrolyte at a finite temperature. Theoretically, the approach suffers from a certain arbitrariness, but it gives quite good results for the case of unscreened Coulomb forces.

The interaction energy for a polyelectrolyte consisting of $N$ monomers with Coulombic repulsion forces is given by

$$E = \sum_{kl} \frac{1}{r_{kl}} \qquad (23)$$

The variational average $\langle E \rangle_V$ is then perfectly convergent. It is however difficult to evaluate, and we will for that reason also here employ the simplifying trick of eq. (11). In this case, the resulting non-uniqueness is used to replace $\mathbf{v}_i^2$ by 1 everywhere; thus, $r_{kl}$ will evaluate to

$$r_{kl} = \sqrt{|k - l| + \sum_{i \neq j} \mathbf{v}_i \cdot \mathbf{v}_j} \qquad (24)$$

where $i, j$ are restricted to the set of bonds linking monomers $k$ and $l$.

Then it is not a priori clear to what degree the results will make sense for a finite $T$. However, the replacement $\mathbf{v}_i^2 \to 1$ gives the correct result for $\langle r_{kl}^2 \rangle_V$, and we argue that it should be a fair approximation also for $\langle 1/r_{kl} \rangle_V$ — it is certainly correct for $T \to 0$ ($|\mathbf{v}_i| \to 1$), and qualitatively correct for large $N$ when $T \to \infty$ ($\mathbf{v}_i \to 0$).

Thus armed with some confidence, we have used this approach to generate configurations in terms of $\{\mathbf{v}_i\}$ for polyelectrolytes ranging in size from 20 to 1024, by iterating equations analogous to eqs. (21,22). We have chosen to characterize the configurations by their r.m.s. end-to-end distance $r_{ee}$, which in the variational approach is given by

$$r_{ee}^2 = (N - 1) + \sum_{i \neq j} \mathbf{v}_i \cdot \mathbf{v}_j \qquad (25)$$

As can be seen from table 4, the results are in surprisingly good agreement with Monte Carlo (**MC**) data.

We have also attempted a similar comparison for polymers with a screened Coulomb interaction. The results in that case are not nearly as good as in the unscreened case, and it is not difficult to see why. With screening, the interaction turns short-range, and for large molecules, essentially all the rotors will be identical in size and direction, $\mathbf{v}_i = \mathbf{v}$, with $v$ independent of $N$. Thus, we get for the end-to-end distance

$$r_{ee}^2 \approx (N-1) + \frac{(N-1)(N-2)}{2}\mathbf{v}^2 \qquad (26)$$

which goes like $N^2$ (unless $\mathbf{v} = 0$, which happens for high enough $T$). Thus, $r_{ee}$ will scale like $N$, which is clearly unphysical for a short range potential. This failure is probably more due to the linear Ansatz for $E_V$ being unsuitable for this problem, and not so much to the crudeness of the additional approximation of eq. (11).

| $N$  | Var.  | MC    | % dev |
|------|-------|-------|-------|
| 20   | 11.53 | 12.06 | -4.6  |
| 40   | 26.80 | 26.43 | 1.3   |
| 80   | 59.05 | 57.07 | 3.5   |
| 160  | 125.8 | 122.1 | 3.0   |
| 256  | 207.6 | 202.5 | 2.5   |
| 512  | 429.5 | 422.0 | 1.4   |
| 1024 | 880.3 | 870.2 | 1.1   |

Table 4: End-to-end distance $r_{ee}$ for polyelectrolytes. Comparison of variational results with MC data for different $N$ for $T = 0.837808$ (room temperature). The last column lists relative deviations between the variational method and MC results in percentage. The errors in the MC data are approximately 0.1 %.

## 7 Summary

We have developed a variational approach for finding approximate energy minima for proteins modelled by polymers with Lennard-Jones pair interactions between pointlike monomers. It has been numerically explored for two cases — dense-packed systems and more elongated ones.

In the latter case, phenomenological pair potentials were used with random amino-acid sequences based on natural frequencies of occurrence. This leads to effective coupling distributions far more narrow than those commonly assumed in generic investigations of spectra and stability issues [9, 8].

For dense-packed systems the variational performance compares favourably with a gradient descent method with respect to solution quality, whereas for the more elongated ones the gain is very small. We interpret this difference as being due to the fact that the elongated chain provides less of an optimization challenge as compared to the dense packed ones. Hence there is a less pronounced difference between the methods.

The deterministic variational method fails to find in a consistent way the low energy states produced by a stochastic simulated annealing algorithm. However, the latter requires a factor 100 more in CPU consumption.

As a by-product we have devised a regularization of the Lennard-Jones potential, where the short-distance steepness is gradually turned on in the updating process, thereby avoiding ill-behaved dynamics. Most deterministic (and some stochastic) methods will benefit from such a regularization.

The variational method has been applied also to the case of polyelectrolytes at non-zero temperature, where the procedure is somewhat more ambiguous. For the case of unscreened repulsions the method yields good results as judged by MC, whereas it breaks down for the screened case.

## Acknowledgements

We thank Anders Irbäck for aiding us with the choice of Lennard-Jones couplings and Bo Jönsson for providing us with polyelectrolyte MC data.